\documentclass[12pt]{article}
\usepackage{amsmath}
\usepackage[varg]{txfonts}
\usepackage{graphics,graphicx,color}  
\newcommand{\lan}{\langle}
\newcommand{\rrr}{\rangle}
\newcommand{\be}{\begin{equation}}
\newcommand{\ee}{\end{equation}}

\def\beq{\begin{eqnarray}}
\def\eeq{\end{eqnarray}}
\def\beqs{\begin{eqnarray*}}
\def\eeqs{\end{eqnarray*}}

\begin{document}

\begin{titlepage}

\begin{center}

{\Large \bf Archimedes Force on Casimir Apparatus }

\bigskip


{V.Shevchenko$^{1,2}$, E.Shevrin$^1$}

\bigskip

{\it  $^1$National Research Centre "Kurchatov Institute", \\
ac. Kurchatova square, 1, Moscow 123182 Russia}
\\
\medskip
{\it $^2$Far Eastern Federal University, \\
Sukhanova street, 8, Vladivostok 690950 Russia \\
}
\end{center}

\bigskip

\begin{abstract}
This letter addresses a problem of Casimir apparatus in dense medium, put in weak gravitational field. The falling of the apparatus has to be governed by the equivalence principle, with proper account for contributions to the weight of the apparatus from its material part and from distorted quantum fields. We discuss general expression for the corresponding force in metric with cylindrical symmetry. By way of example we compute explicit expression for Archimedes force, acting on the Casimir apparatus of finite size, immersed into thermal bath of free scalar field. It is shown that besides universal term, proportional to the volume of the apparatus, there are non-universal quantum corrections, depending on the boundary conditions.
\end{abstract}

\end{titlepage}

\section{Introduction}
\label{intro}

Reflections about matter dynamics in gravitational field are among the most fruitful themes in the history of physics. Well known legends about Newton, inspired by the falling apple in his mother's garden, or about Galileo, dropping balls from the top of the Leaning Tower of Pisa are good examples. In modern times, classical tests of General Relativity such as light deflection by the Sun's gravity; weighting-the-photon experiments of Pound-Rebka type; Shapiro delay; neutron interferometry in gravitational field; ALPHA, AEGIS and GBAR experiments at CERN, exploring falling antimatter, continue the same line of studies.

Of particular interest are quantum field theoretic physics in classical gravitational field, where Hawking radiation is the best known phenomenon \cite{8}.
Needless to say that the problem of genuine gravitational interaction between parts of intrinsically quantum object (for example, between two entangled photons) cannot be addressed in semiclassical approach, leaving aside the fact that it is beyond our current experimental abilities. It is to be stressed that we have no direct experimental information how an elementary particle like proton gravitationally interacts with another one at, say, distances $\sim 10^{-10} $ meters. Therefore naive extrapolation of Newton gravity law to the Planck distances $\sim 10^{-35}$ meters could be plainly wrong, as various extra dimensions scenarios suggest. In other words, the "ultimate" ultraviolet fundamental scale can well have nothing to do with the conventional Planck distance (calculated from long-distance asymptotic of the gravitational interaction, described by the Newton constant $G$).

Coming back to the case when semiclassical treatment is appropriate, the simplest example is non-relativistic motion of a test body in external weak gravitational field. The basic fact governing this type of motion is well known from school textbooks: the force acting on the body is proportional to its mass and directed along the free fall acceleration:
\be
{\bf f} = m {\bf g} = \rho V {\bf g}
\label{e1}
\ee
where $\rho = m/V$ is average density and $V$ is the body's volume. Simplicity of this formula should not camouflage a highly nontrivial fact, that the force depends on the only parameter of the  body - its mass (and not, for example, on its chemical composition, entropy etc). Combined with the Newton's second law of motion this fact has, of course, direct relation to the celebrated equivalence principle.

The situation gets more complex if the test body is immersed into gas or fluid. The expression (\ref{e1}) is to be replaced in this case by
\be
{\bf f} = (\rho - \rho_f) V {\bf g}
\label{e2}
\ee
where $\rho_f$ is the fluid's density, and the term proportional to $\rho_f$ is known as Archi\-me\-des force. The expression (\ref{e2}) hides a few approximations and there are a few relevant small parameters. First, the independence of this force on any characteristic features of the body other than its volume is by no means trivial. It is based on smallness of a ratio of gas/fluid molecules size to that of the body (and also holes in the body's surface etc), which makes continuous medium approximation applicable.\footnote{This is just what helped Archimedes to find out the volume of King's Hiero crown in well known legend.} Another parameter is the Planck constant $\hbar$ - the result (\ref{e2}) is of course purely classical and may get quantum corrections, for example, if typical quantum correlation length in the fluid is
comparable with the body size. Also needless to say that (\ref{e2}) is valid in non-relativistic and weak gravitational field approximations.
Last but not least, the expression (\ref{e2}) is invariant under shifts $\rho \to \rho + \mbox{const}$. It is "self-renormalized" in this sense and piece of vacuum (or any other medium in stationary case) with the "mass"
\be
\frac{1}{c^2} \int dV \> \lan T^{00}\rrr
\ee
does not "fall" in external gravitational field, because there is compensating "pressure" on this piece of exactly the same magnitude from surroundings, directed "upwards".

This letter analyzes weighting of the Casimir apparatus in weak gravitational field. The problem has attracted some attention in recent years
\cite{10,11,12,13,14,15,16,17,18} and there used to be controversy in the literature we will mention below. We argue that the key point is physically correct definition of the weighting procedure, since there is no possibility to weight Casimir energy alone - one always measure the weight of Casimir apparatus as a whole. The weighting procedure and the results are to be universal and applicable to any Casimir apparatus, not only to two parallel plate Casimir cavity, usually taken as example. Our aim is to discuss such procedure and to apply it to concrete case of Casimir cavity in thermal bath of massless scalar field.

\section{Archimedes Force}
\label{sec-1}

The basic ingredient is quantum field theoretical average of energy-momentum tensor $\lan T_{\mu\nu} (x) \rrr $, where average over fields is computed with %
the standard  integration measure ${\cal D} \Phi$, normalized to have  $\lan  1  \rrr = 1$.
In geometric setup used by us in this paper,\footnote{By the word "geometric" we mean neglect of dynamical properties of the boundaries like frequency-dependent reflectivity etc.} Casimir apparatus is encoded by some $x$-dependent measure deformation, ${\cal D} \Phi \to {\cal D}' \Phi$, corresponding to constraints the fields have to obey on the boundary or in interior of the apparatus.  For example, in classical Casimir setup of two infinite ideally conducting parallel planes interacting with electromagnetic field this deformation looks like\cite{bordag}
\be
{\cal D} A_{\rho}(x) \to {\cal D} A_{\rho}(x) \> \delta({\tilde F}^{\mu 3}(x_3=a_1)) \delta({\tilde F}^{\mu 3}(x_3=a_2))
\ee
where ${\tilde F}^{\mu\nu}$ is dual field strength, $a_1$, $a_2$ - coordinates of the planes along the third axis (the axes $1$ and $2$ are in the planes). The form of the above expression is quite general and in many cases one can write  ${\cal D}' \Phi =  {\cal D} \Phi \> \Delta[\Phi]$ with some functional of the fields. One could think of Casimir plates as of "passive detectors". In other words, one selects only those field configurations, where parallel electric and normal magnetic fields at the position of the plates are zero at all moments of time. This is like having quantum or classical particle in a corridor made of hard walls so that any subsequent measurement of the particle coordinate will definitely return a result inside the corridor. Uncertainty relation causes pressure on the boundaries (walls and plates in the examples above), which depends, in particular, on how hard (conductive) they are.

Coming back to the Casimir apparatus in weak gravitational field, writing the metric as $g_{\mu\nu} = \eta_{\mu\nu} + 2 h_{\mu\nu}$, one gets for the energy shift at the leading order of semiclassical approximation \cite{12}
\be
\delta E_g = -\int d^3 x \> h_{\mu\nu}(x) \lan T^{\mu\nu} \rrr(x)
\label{io8}
\ee
In geometric setup the average $\lan T^{\mu\nu} \rrr(x)$ has two parts - the "material" one, corresponding to the objects (planes, cavities, robes, springs etc) the Casimir apparatus is made of, and the "field" part.
For nonrelativistic case spatial components of the energy-momentum tensor are suppressed by inverse powers of the speed of light, $\left| T^{00} \right| \gg \left| T^{ij} \right|$ and only temporal component of the metric tensor $h_{00}(x)$ is relevant. This can be correct approximation for the material part of $\lan T^{\mu\nu} \rrr(x)$, but certainly not for its "field" part we are interested in here. Consequently it is easy to check, that various choices of the metric $h_{\mu\nu}$ lead to different answers for $\delta E_g$, even if all these choices correspond to uniform field with free fall acceleration ${\bf g}$.
Moreover, the energy becomes orientation-dependent for some choices, in gross contradiction with the equivalence principle and scalar nature of mass. This is physically unacceptable and should be resolved.

The source of the problem was identified in \cite{12} as gauge non-invariance of (\ref{io8}). Indeed, (\ref{io8}) is invariant under weak field gauge transformation $h_{\mu\nu} \rightarrow h_{\mu\nu} + \partial_\mu \xi_\nu + \partial_\nu \xi_\mu$ only if $\partial_\mu T^{\mu\nu} = 0$.
The energy-momentum tensor is covariantly conserved for the combined "material + field" system
\be
\nabla_\mu T^{\mu\nu} = 0
\label{emt}
\ee
but not for the "field" part alone. Thus two logically possible alternative paths can be chosen: either one is to include the material part and carefully work with the full energy-momentum tensor, obeying (\ref{emt}), or one is to argue, that this or that choice of the metric is more physical than another choices and compute the force using the distinguished metric. Mostly the latter path was followed in the literature with the motivation for preferable role of Fermi metric choice $c^2 h_{00} = {\bf g}{\bf z} \;\; ; \;\; h_{ij} = 0$ and the result for the weight of Casimir energy (in classical two plates case)
\be
{\bf f} = {\bf g} \>\frac{E_C}{c^2}\> S \;\;\; \mbox{where} \;\;\; E_C = - \frac{\pi^2 \hbar c}{720 a^3}
\label{io9}
\ee
and $S$ stays for the plates area. The energy-momentum tensor of the system is given by $\lan T^{\mu\nu}\rrr = (E_C / a)\times \mbox{diag}(1,-1,-1,3)$ between the plates and zero outside \cite{18}.

Thus, the cavity feels small upward push and Casimir energy gravitates as any other form of energy in accord with the equivalence principle. In a sense, the answer (\ref{io9}) could have been written without any computations, if $E_C$ is known. However, the arguments based on physically distinguishable role of a particular parametrization of the metric are difficult if not impossible to generalize to other cases. In particular, it is not clear how to write the next order ${\cal O}( h_{\mu\nu} h_{\rho\sigma})$ correction to (\ref{io9}). It is also important, that the result (\ref{io9}) by its nature should be quite general, which, however, is also not clear taking into account that the methods used for its derivation (see e.g. \cite{11}) heavily use properties of particular two infinite parallel planes geometry. This calls for systematic derivation applicable beyond the weak field approximation and for arbitrary Casimir apparatus.

We argue here that weighting methodology suggested in \cite{19} can be naturally adopted to the Casimir apparatus weighting problem. Consider static metric of the following form
\be
ds^2 = g_{00}(x_3)c^2dt^2 + g_{33}(x_3) d x_3^2 + d {\bf x}_\bot^2
\label{s}
\ee
with the choice of $x_3$-coordinate axes such that $g_{00}(x_3=0) = -1$ (we denote $x_3$ coordinate as $z$ below for simplicity of notation).
We are to weight, following the symmetry of the above choice, two large identical boxes with some identical boundary conditions for the fields on their internal boundaries (see Figure 1). We put inside the box number 2 the Casimir apparatus of much smaller size, i.e. we assume a set of conditions encoded by some functional
\be
{\cal B}[\phi(x)] = 0 \;\; \mbox{for} \;\; x\in V
\label{io11}
\ee
put on the fields inside the volume of the apparatus $V$ or, in particular case, only on its boundary $S = \partial V$. There is no apparatus in the box number 1.

\begin{figure}
\noindent\centering{
\includegraphics[width=75mm]{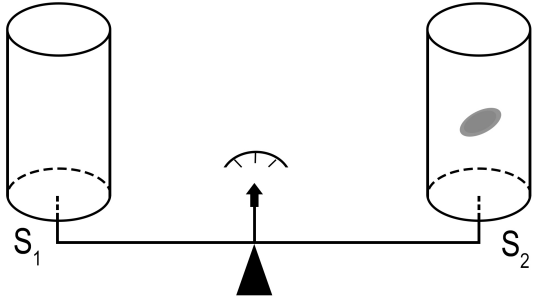}
}
\caption{Weighing of the Casimir apparatus (shown as dark ellipse in the right box).}
\label{figCurves}
\end{figure}

Then, following \cite{19} we consider a function:
\be
w_k(z) = \int\limits_{S_k} d^2{\bf x}_\bot \> \sqrt{-g_{00}} \cdot g_{33} \lan T^{33}\rrr_k
\label{ki0}
\ee
The index $k=1,2$ is the box label and integration goes over section of the boxes at constant $z$ (sections are assumed to be of arbitrary shape, but $z$-independent, i.e. geometry is cylindrical). The measures $\lan ... \rrr_k$ take into account Casimir apparatus conditions (\ref{io11}) in the box 2. For Minkowskii space with the metric $g_{\mu\nu} = \eta_{\mu\nu} = \mbox{diag} (-1,1,1,1)$ the function (\ref{ki0}) is nothing but the integrated pressure, i.e. for $z=0$ it is the force acting on the bottom plane of the box. It is obvious that all contributions to this force from "material" parts of the boxes 1 and 2 are identical by construction.

The difference of these integrated pressures at $z=0$ we call, by definition, the weight of our Casimir apparatus:
\be
f = w_2(0)-w_1(0)
\label{p}
\ee
To express this force in terms of energy-momentum tensor integrals,
we are to take into account that the total energy-momentum tensor is covariantly conserved inside each box. For the metric choice (\ref{s})
the equation (\ref{emt}) reads
\be
\frac{\partial w(z)}{\partial z}  = \frac{\partial  \sqrt{-g_{00}}}{\partial z} \int\limits_{S_k} d^2{\bf x}_\bot \> g_{00} \lan T^{00}\rrr
\label{lab}
\ee
where we have used the definition (\ref{ki0}). Integrating (\ref{lab}) over the entire boxes and assuming  $\lim\limits_{z\to\infty} [{\lan T^{33}\rrr}_2(z) - {\lan T^{33}\rrr}_1(z) ] = 0$ (which physically corresponds to finiteness of the Casimir apparatus), we obtain the following final answer
\be
f = \frac12 \int d^3{\bf x}\> \sqrt{-g_{00}(z)}\left(\frac{\partial g_{00}(z) }{\partial z}\right) \left[{\lan T^{00}\rrr}_2(x) - {\lan T^{00}\rrr}_1(x)\right]
\label{f}
\ee
We discuss this result in the next section.

\section{Discussion and applications}

The most important property of the expression (\ref{f}) is its independence on spatial components of the metric and energy-momentum tensors.
To get that the particular form (\ref{s}) of the metric tensor was crucial. It is clear, in particular, that the weighting procedure we use would not be operational for the case of transverse coordinates $x_1 , x_2$ - dependent metric tensor.
On the other hand, one has no need to take $g_{33}=1$ and nowhere we have used weak field approximation. Therefore in is legitimate to expand (\ref{f}) over difference $(-1 - g_{00})$. The leading term corresponds to the standard choice $g_{00} = -1 + 2{\bf gz}/c^2 + {\cal O}(1/c^4)$:
\be
{\bf f_0} = {\bf g} \frac{1}{c^2}  \int d^3{\bf x}\> \left[{\lan T^{00}\rrr}_2(x) - {\lan T^{00}\rrr}_1(x)\right]
\label{fmain}
\ee
For Casimir plates discussed above equation (\ref{fmain}) reproduces the result (\ref{io9}). On the other hand, in classical limit we come back to (\ref{e2})  taking into account that
\be
\int\limits_{V_{body}} d^3{\bf x}\> {\lan T^{00}\rrr}_2(x) = m c^2 \;\; ; \;\; \int\limits_{V_{body}} d^3{\bf x}\>  {\lan T^{00}\rrr}_1(x) = \rho_f c^2 V_{body}
\ee
and $\lan T^{00}\rrr_2(x) = {\lan T^{00}\rrr}_1(x)$ for $x$ outside the body. Needless to say that this last condition does not take place for quantum "field part" of the total energy-momentum tensor, since the body distorts fields around it and its energy is delocalized in this sense.

The important issue is UV-divergencies of (\ref{f}). Since the force is physical observable, we expect they all get cancelled. The detailed picture of such renormalization can be rather tricky, as examples \cite{10,14} clearly show. We have to take care only about divergencies, related to the Casimir apparatus. General intuition suggests that the former ones renormalize the nonrelativistic mass the the material objects the apparatus is made of. From this point of view, while distinction between "body" and "medium" is clear in non-relativistic and non-quantum limits, it is somehow lost in general case since the body in question - Casimir apparatus - is surrounded by the cloud of quantum fields, distorted by its presence and this distortion contributes to its total rest mass. Contrary to classical vacuum, which at least in principle can be cleaned to any desired level, allowing independent measurement of each contribution, one cannot "clean" quantum vacuum by eliminating fluctuating quantum fields out of it.

Let us also make a comment on next-to-leading corrections to (\ref{fmain}). They come from two places: expansion of metric-dependent multiplier in (\ref{f}) and expansion of energy-momentum tensor average. It is convenient to rewrite (\ref{f}) as
\be
f = 2 \int d^3{\bf x}\> \left(\frac{\partial h_{00}(z) }{\partial z}\right) \frac{1}{\sqrt{g_{33}(z)}} \> \frac{\delta (W_2 - W_1)}{\delta g_{00}(z)}
\ee
where $W_k$ is the corresponding effective action and the standard definition
\be
\lan T^{\mu\nu}(x) \rrr = \frac{2}{\sqrt{-g}}\frac{\delta W}{\delta g_{\mu\nu}(x)}
\ee
 was used. Next-to-leading correction has the following form:
$$
f = f_0 -  \int d^3{\bf x}\> \left(\frac{\partial h_{00}(z) }{\partial z}\right) h_{33}(z) \left[\lan T^{00}\rrr_2(x) - {\lan T^{00}\rrr}_1(x)\right] +
$$
\be
+ 4 \int d^3{\bf x}\> \left(\frac{\partial h_{00}(z) }{\partial z}\right) \int d^4 x' \> h_{\alpha\beta}(x') \left. \frac{\delta}{\delta  g_{\alpha\beta}(x') }\>\frac{\delta (W_2 - W_1)}{ \delta g_{00}(x)}\right|_{h=0}
\label{fd}
\ee
We see that the dependence on spatial components of the metric appears at the next order. Another correction to the classical Archimedes force - "weight of quantum fluctuations" - corresponds to the last term. It is worth mentioning \cite{20} that in Casimir systems with massless fields fluctuations of energy-momentum tensor components
 \be
 \int d^4 x' \> [ \lan T^{\mu\nu}(x) T^{\alpha \beta}(x') \rrr - \lan T^{\mu\nu}(x) \rrr \cdot \lan T^{\alpha \beta}(x') \rrr ]
 \ee
 are typically not small with respect to average $\lan T^{\mu\nu}(x) \rrr $, so both the second and the third terms in the right hand side of (\ref{fd}) can in general be of the same order.

For general geometry of the Casimir body the right hand side of (\ref{f}) or (\ref{fmain}) is given by some complicated expression, and no universal dependence of the Archimedes force on the body's volume like in (\ref{e2}) can be expected.
To get the latter universality there should be small parameter in the system, as we discussed above. An interesting example is Casimir apparatus in the thermal bath. If geometric size of the apparatus is large compared to thermal wavelength
\be
r \gg \frac{\hbar c}{k_B T}
\label{oj}
\ee
one could think that large temperature expansion is a good approximation. Technically we can realize it using effective action and heat kernel expansion formalism in Euclidean space (see review \cite{21}). It is convenient to start with expression for free energy
\be
\beta F_\beta = -\log \int {\cal{D}} \Phi \> e^{-S[\Phi]}
\ee
where the action for massless minimally coupled free scalar field is given by the standard expression $S=\frac12 \int_0^\beta d\tau \int d^d {\bf x}\>  \Phi \> (-\Box) \> \Phi$ and covariant D'Alembertian is $\Box = \nabla_{\mu}\nabla^{\mu}$.
In Euclidean formalism (see, e.g. \cite{22,23}) one considers theory in $d+1$-dimensional Euclidean space-time with the topology $\mathbb{R}^d \times \mathbb{S}^1$, where length of the latter compact dimension is denoted as $\beta$.
The fields satisfy the conditions of periodicity in Euclidean time
$\Phi \left( {\bf x}, \tau \right) = \Phi \left({\bf x}, \tau + \beta \right)$. Parameter $\beta$ will be associated with inverse temperature in what follows: $\beta = (k_B T)^{-1} $.

As is well known, the temperature-dependent part of one-loop Euclidean effective action can be represented in terms of the corresponding thermal heat kernel $\hat{K}^\beta (s|x,y)$:
\begin{equation}
\beta F_\beta =-\frac{1}{2} \int\limits_0^\infty \frac{ds}{s} \> \left( \mathrm{Tr}  \hat{K}^{\beta}(s) - \mathrm{Tr}  \hat{K}(s) \right)
\end{equation}
where $\hat{K}^{\beta}(s)$ is periodic in Euclidean time solution of the equation
\begin{equation}
\left( \frac{d}{ds} - \Box \right) \hat{K}^{\beta}(s|x,y) = \hat{1}\cdot \delta(s)\delta(x,y)
\end{equation}
with $s$ playing the role of proper time. The temperature dependence of the trace of finite temperature heat kernel can be factorized as \cite{23}
\begin{equation}
\mathrm{Tr}\hat{K}^{\beta}(s) = \frac{\beta}{(4\pi s)^{1/2}} \> \theta_{3}\left(0, e^{-\frac{\beta^2}{4s}} \right) \int d^{d} {\bf x} \> \mathrm{tr} \> \hat{K}_d (s|{\bf x}, {\bf x})
\end{equation}
where $\theta_3(a,b)$ is Jacobi function and we denote $d$-dimensional zero-temperature kernel $\hat{K}^{\infty}(s)$ as $\hat{K}_d(s)$.

The key result \cite{dw1,dw2,dw3,dw4}  is expression for the heat kernel asymptotic expansion on the manifold $\mathcal{M}$ in powers of the proper time:
\begin{equation}
 \int d^{d} {\bf x} \> \mathrm{tr} \> \hat{K}_d (s|{\bf x}, {\bf x}) =  \frac{1}{(4\pi s)^{\frac{d}{2}}}\sum \limits_{n=0}^{\infty}\left( s^{n}A_n+s^{n/2}B_{n/2}\right)
\end{equation}
where the coefficients are given by
\begin{equation}
A_n = \int_{\mathcal{M}}d^{d}x\> \sqrt{-g(x)}\> a_n(x) \;\;\; ; \;\;\;
B_{n/2}=\int_{\mathcal{\partial M}} d^{d-1}x \> \sqrt{\gamma(x)}\> b_{n/2}(x)
\label{ab}
\end{equation}
for integer and half-integer powers. Here $g(x)$ and $\gamma(x)$ denote the determinants of the bulk and induced boundary metrics, respectively. The surface integrals at the boundaries $B_{n/2}$ are build of local invariants incorporating such local characteristics of the surface as its extrinsic curvature $K_{\mu\nu}$ etc.

In the problem under discussion the manifolds $\mathcal{M}$ correspond to intrinsic space inside the boxes. Let us denote as $V_{box}$ the volume of the boxes (identical for the box 1 and the box 2), then by $V$ we denote the volume of Casimir apparatus, located in the box 2. In the same way we denote as $S_{box}$ the identical surface area of the boxes and by $S$ - the surface area of Casimir apparatus. Then taking into account that $a_0 = 1$, one obtains, at the leading order, for the free energy in the box 1 (without Casimir apparatus)
\be
F^{(1)}_\beta = - \frac{\pi^2}{90} \frac{1}{\beta^4} V_{box} + b \frac{\zeta(3)}{8\pi}\frac{1}{\beta^3} S_{box} + {\cal O}(\beta^{-2})
\ee
while for the box 2, excluding the apparatus:
\be
F^{(2,out)}_\beta = - \frac{\pi^2}{90} \frac{1}{\beta^4} ( V_{box} - V)  + b \frac{\zeta(3)}{8\pi}\frac{1}{\beta^3} (S_{box} + S) + {\cal O}(\beta^{-2})
\ee
The parameter $b$, proportional to $b_{1/2}$ from (\ref{ab}), encodes boundary conditions for the field, for particular case of Dirichlet boundary conditions $b=1$, while $b=-1$ for Neumann ones \cite{dw2,dw4}. In principle, one can consider the case (quasistationary for small thermal conductivity of the material the apparatus is made of) with different temperatures inside and outside the apparatus and add contribution to the total free energy in the box 2 from the internal volume of the apparatus:
\be
F^{(2,in)}_\beta = - \frac{\pi^2}{90} \frac{1}{{\beta}_{in}^4}  V  + b \frac{\zeta(3)}{8\pi}\frac{1}{{\beta}_{in}^3} S + {\cal O}(\beta^{-2})
\ee
Then, using the relation
\be
U = F_\beta - T \frac{\partial F_\beta}{\partial \beta}
\ee
for internal energy, we get for the leading and first sub-leading contributions to Archimedes force (\ref{fmain})
\be
{\bf f} - m{\bf g} = {\bf g} \frac{\hbar }{c} \left( \frac{\pi^2 }{30} \left(T_{in}^4 - T_{out}^4 \right) \left( \frac{k_B}{\hbar c} \right)^4 \cdot V -  b \frac{\zeta(3)}{4\pi} \left(T_{in}^3 + T_{out}^3 \right)  \left( \frac{k_B}{\hbar c} \right)^3 \cdot S \right)
\label{y}
\ee
This result is worth commenting. First, as it should be by design of our weighting procedure, all factors depending on geometry of the boxes have been cancelled. The expansion goes in parameter $(\hbar c / k_B T)(S / V)$ which is assumed to be small\footnote{The parameter $\hbar c / k_B T = 7.6\cdot 10^{-6} $ meters at 300 K.} according to condition (\ref{oj}). The first term in the right hand side is nothing but the weight of thermal photon gas.\footnote{This physics described by (\ref{y}) should not be misinterpreted as physics of flying balloons with heated air inside, where key factor is pressure gradient dependence on temperature.}
This term is universal and scales as volume. The next-to-leading non-universal term depends on boundary conditions and scales as area. Due to scalar nature of the problem both inside and outside parts of the boundary contribute with the same sign and this term does not vanishes but doubles for equal inside and outside temperatures.
The importance of such terms and surface-dependent effects they describe for various metrology problems like precise calibration of thermometers etc was stressed in \cite{23}.

\section{Conclusions}

We discussed universal expression for the Archimedes force on Casimir apparatus (\ref{f}) applicable to any kind of the latter. It is shown that only zeroth component of the energy-momentum tensor contributes to it. In case of Casimir apparatus in the thermal bath of massless scalar field next-to-leading correction of quantum origin (\ref{y}) is computed.

The effects discussed above are extremely tiny. Indeed, the ratio between "mass" part of the force (equal to $|m{\bf g}|$) and Archimedes part (given by the right hand side of (\ref{y})) is of the order of $10^{20}$ for a body of macroscopic mass and size at room temperature. Direct detection of such a small static force seems to be hopeless. It is worth remembering nevertheless examples in the history of physics when extremely weak effects became detectable with the help of amplifiers like multiplicity, interference or resonance. Huge value of Avogadro number allowing with multi-ton detectors to reach limits on the lifetime of a proton far exceeding the age of the Universe, recent observation of gravitational waves \cite{gw}, interesting suggestions to use sophisticated balances to weight internal energy \cite{calloni1,calloni2} are good examples demonstrating power of these techniques, respectively. In this respect, the problem to find experimentally reasonable "amplifying factor" for weak gravity of quantum states/energies is not closed and in our opinion still worth studying. This land is to large extent {\it terra incognita} experimentally and one can hope for surprises here.

\section{Acknowledgements}

One of the authors (V.Sh.) thanks Yu. Gusev for discussions. The work was partly supported by the RFBR Grant 14-22-03030.


\begin{thebibliography}{99}
\bibitem{8}
S.~W.~Hawking,
  Commun.\ Math.\ Phys.\  {\bf 43},  199 (1975)
   Erratum: [Commun.\ Math.\ Phys.\  {\bf 46}, 206 (1976)].
\bibitem{10}
  K.~A.~Milton, K.~V.~Shajesh, S.~A.~Fulling and P.~Parashar,
  Phys.\ Rev.\ D {\bf 89} 064027 (2014)
\bibitem{11}
  K.~V.~Shajesh, K.~A.~Milton, P.~Parashar and J.~A.~Wagner,
  J.\ Phys.\ A {\bf 41} 164058 (2008)
 \bibitem{12}
  S.~A.~Fulling, K.~A.~Milton, P.~Parashar, A.~Romeo, K.~V.~Shajesh and J.~Wagner,
  Phys.\ Rev.\ D {\bf 76} 025004 (2007)
 \bibitem{13}
  K.~A.~Milton, P.~Parashar, K.~V.~Shajesh and J.~Wagner,
  J.\ Phys.\ A {\bf 40}  10935 (2007)
 \bibitem{14}
   K.~A.~Milton,
  Lect.\ Notes Phys.\  {\bf 834} 39 (2011)
\bibitem{15}
G.~Bimonte, E.~Calloni, G.~Esposito, G.~M.~Napolitano and L.~Rosa,
  arXiv:0904.4864 [hep-th].
\bibitem{16}
 G.~Bimonte, E.~Calloni, G.~Esposito and L.~Rosa,
  Phys.\ Rev.\ D {\bf 76}  025008 (2007)
\bibitem{17}
  G.~Bimonte, E.~Calloni, G.~Esposito and L.~Rosa,
  Phys.\ Rev.\ D {\bf 74} 085011 (2006);
   [Phys.\ Rev.\ D {\bf 75}  049904 (2007)]
   [Phys.\ Rev.\ D {\bf 75} 089901 (2007)]
   [Phys.\ Rev.\ D {\bf 77} 109903 (2008)]
\bibitem{18}
 L.~S.~Brown and G.~J.~Maclay,
  Phys.\ Rev.\  {\bf 184}  1272 (1969).
\bibitem{bordag}
M.~Bordag, D.~Robaschik and E.~Wieczorek,
  Annals Phys.\  {\bf 165}, 192 (1985).
\bibitem{19}
T.Padmanabhan, {\it Gravitation: Foundations and Fronties,} Cambridge University Press, 2010.
\bibitem{20}
V.~Sopova and L.~H.~Ford,
  Phys.\ Rev.\ D {\bf 66}  045026 (2002)
\bibitem{21}
D.~V.~Vassilevich,
  Phys.\ Rept.\  {\bf 388} 279 (2003)
  \bibitem{22}
   Y.~Gusev and A.~Zelnikov,
  Class.\ Quant.\ Grav.\  {\bf 15} 13 (1998)
 \bibitem{23}
 Y.~V.~Gusev,
  Russ.\ J.\ Math.\ Phys.\  {\bf 22} 9 (2015)
  \bibitem{24}
  Y.V.Gusev, Russ. J. Math. Phys. {\bf 23} 1 (2016)
 \bibitem{dw1}
B. S. DeWitt., Phys.Rev. {\bf 162} 1195 (1967).
\bibitem{dw2}
L. Dolan and R. Jackiw., Phys. Rev. {\bf D9} 3320 (1974).
\bibitem{dw3}
J. S. Dowker and G. Kennedy., J. Phys. A: Math. Gen. {\bf 11} 895 (1978).
\bibitem{dw4}
I. G. Avramidi., Physics of Atomic Nuclei {\bf 56} 138 (1993).
 \bibitem{gw}
   B.~P.~Abbott {\it et al.} [LIGO Scientific and Virgo Collaborations],
   Phys.\ Rev.\ Lett.\  {\bf 116}, 061102 (2016)
  [arXiv:1602.03837 [gr-qc]].
\bibitem{calloni1}
  E.~Calloni {\it et al.},
    Phys.\ Rev.\ D {\bf 90} 022002 (2014)
\bibitem{calloni2}
  E.~Calloni {\it et al.},
  arXiv:1511.04269 [gr-qc].
 \end{thebibliography}
\end{document}